\documentclass[journal]{IEEEtran}

\usepackage{amsmath,amssymb,mathtools,bm}
\usepackage{graphicx}
\usepackage{booktabs,multirow}
\usepackage{subfig}
\usepackage[ruled,vlined]{algorithm2e} 
\usepackage{hyperref}
\hypersetup{hidelinks}

\title{Carbon-Aware Orchestration of Integrated Satellite Aerial Terrestrial Networks via Digital Twin}

\author{Shumaila  Javaid, and Nasir Saeed,~\IEEEmembership{Senior Member,~IEEE}
   \thanks{S. Javaid is with the College of Electronics and Information Engineering, Tongji University, Shanghai 201804, and the State Key Laboratory of Autonomous Intelligent Unmanned Systems, Tongji University, Shanghai 201210, China\\
    N. Saeed is with the Department of Electrical and Communication Engineering, College of Engineering, UAE University, Al-Ain 15551, UAE (e-mail: mr.nasir.saeed@ieee.org).}}

\begin{document}
\maketitle

\begin{abstract}
 Integrated Satellite–Aerial–Terrestrial Networks (ISATNs) are envisioned as key enablers of 6G, providing global connectivity for applications such as autonomous transportation, Industrial IoT, and disaster response. Their large-scale deployment, however, risks unsustainable energy use and carbon emissions. This work advances prior energy-aware studies by proposing a carbon-aware orchestration framework for ISATNs that leverages Digital Twin (DT) technology. The framework adopts grams of CO$_2$-equivalent per bit (gCO$_2$/bit) as a primary sustainability metric and implements a multi-timescale Plan–Do–Check–Act (PDCA) loop that combines day-ahead forecasting with real-time adaptive optimization. ISATN-specific control knobs, including carbon-aware handovers, UAV duty-cycling, and renewable-aware edge placement, are exploited to reduce emissions. Simulation results with real carbon intensity data show up to 29\% lower gCO$_2$/bit than QoS-only orchestration, while improving renewable utilization and resilience under adverse events.
\end{abstract}

\begin{IEEEkeywords}
6G, net-zero, digital twin, ISATN, carbon-aware networking, O-RAN, MPC, reinforcement learning.
\end{IEEEkeywords}

\

\section{Introduction}
The rapid evolution of next-generation communication systems is driving the integration of Satellite, Aerial, and Terrestrial Networks (ISATNs) into a unified infrastructure capable of delivering seamless global connectivity. This convergence is critical for enabling emerging applications such as autonomous transportation, Industrial Internet of Things (IIoT), remote healthcare, and disaster response, where reliable, low-latency, and high-capacity communication is essential \cite{zhu2021integrated}. However, the energy consumption associated with operating dense terrestrial base stations, satellite constellations, and aerial platforms introduces significant carbon emissions, posing new challenges for designing energy-efficient and environmentally sustainable integrated networks.

As communication networks scale toward 6G and beyond, addressing carbon emissions and energy optimization has become a priority. The increasing reliance on renewable energy sources and fluctuating carbon intensity in power grids demand intelligent orchestration mechanisms capable of balancing Quality of Service (QoS) with environmental impact. Traditional optimization approaches focus mainly on performance metrics such as throughput, latency, or coverage, often neglecting the environmental cost. Moreover, the highly dynamic nature of ISATNs, with varying weather conditions, orbital parameters, and aerial mobility patterns, further complicates network management, making static or reactive solutions inadequate \cite{geraci2022integrating}.

Digital Twins (DT) technology has emerged as a promising paradigm, providing a continuously synchronized replica of the network that integrates telemetry, forecasts demand and energy conditions, and evaluates ``what-if'' scenarios. DTs thus provide a foundation for proactive and carbon-aware orchestration in ISATNs.
Despite their promise, most existing research on sustainability in communication networks has remained domain-specific rather than holistic. Prior efforts have targeted energy efficiency within individual domains of the network. Terrestrial systems have investigated energy-efficient routing, renewable-powered base stations, and AI-based traffic optimization, while satellites have focused on resource allocation and constellation management, and UAV networks on trajectory optimization and energy harvesting \cite{thantharate2024greensky,javaid2024leveraging}. However, these efforts remain limited to isolated segments and largely overlook real-time carbon intensity metrics or DT-based orchestration across multiple layers. This gap highlights the need for a holistic, carbon-aware orchestration framework supported by predictive modeling and real-time adaptability.

This work proposes a carbon-aware orchestration framework that integrates environmental data into network decision-making. By leveraging real-time measurements of carbon intensity, renewable availability, and power consumption, orchestration algorithms can dynamically adjust traffic routing, compute allocation, and spectrum management to minimize environmental impact. The framework adopts \mbox{gCO$_2$e/bit} (grams of CO$_2$-equivalent per delivered bit) as the primary sustainability metric for evaluating network performance. Achieving reductions in this metric requires advanced modeling and predictive capabilities that operate across multiple layers and domains. DTs can play a central role by providing a high-fidelity virtual replica of the physical network to simulate scenarios, optimize configurations, and support proactive decision-making. The combination of carbon-aware intelligence and DTs enables operators to forecast energy demand, evaluate sustainability strategies, and execute orchestration actions at scale. The main contributions are as follows:


\begin{itemize}

\item First, we formalize \mbox{gCO$_2$e/bit} as a primary sustainability metric alongside QoS (throughput, latency, availability) and formulate a sustainability-constrained orchestration problem for 6G networks, targeting higher performance with significantly reduced carbon emissions per bit.

\item Then, we design a DT architecture that implements a Plan–Do–Check–Act (PDCA) control loop. The DT forecasts traffic patterns, mobility, and grid carbon intensity in the planning stage. It applies routing, placement, and scheduling policies during execution, evaluates QoS, energy, and carbon outcomes in the checking stage, and re-optimizes configurations in real time to enable carbon-aware operation.

 \item Finally, we identify and exploit control knobs unique to ISATNs. These include carbon-aware satellite handover scheduling, UAV duty-cycling and sleep policies, and edge-service placement co-optimized with renewable availability. Together, these mechanisms jointly improve QoS and reduce \mbox{gCO$_2$e/bit}.

\end{itemize}

By systematically combining these elements, our work establishes a holistic architecture and methodology for building greener, more resilient, and energy-aware networks in the 6G era.

\section{Why Net-Zero 6G Needs a Digital Twin (DT)}
The transition to 6G networks coincides with growing global efforts to achieve net-zero GreenHouse Gas (GHG) emissions. The information and communications technology (ICT) sector is projected to consume a significant share of global electricity by 2030, with estimates suggesting it may account for over 20\% of worldwide demand if left unmanaged \cite{ericsson2020}. This trend underscores the urgent need for communication systems that deliver higher throughput and lower latency while drastically reducing energy consumption and associated carbon emissions. Although renewable energy penetration is increasing, grid carbon intensity remains highly variable across time and geography due to the fluctuating availability of solar and wind power, storage constraints, and regional energy generation profiles. In this dynamic environment, static network management policies, designed primarily to optimize throughput or latency, are insufficient for minimizing the carbon footprint of large-scale ISATNs.
DTs provide a systematic way to address these challenges by creating a continuously synchronized, high-fidelity virtual replica of the network. The DT integrates telemetry data from base stations, satellites, aerial platforms, and core network elements to construct a holistic view of the system. It forecasts demand patterns, channel states, mobility dynamics, and power grid signals, enabling operators to anticipate carbon intensity fluctuations and proactively optimize network performance. Unlike static planning tools, a DT can simulate ``what-if'' scenarios, evaluate carbon-aware policies, and recommend actionable strategies such as energy-efficient routing, UAV duty-cycling, satellite handover scheduling, and edge computing placement. By continuously monitoring both QoS metrics and carbon Key Performance Indicators (KPIs), the DT serves as a coordination brain for sustainable orchestration across heterogeneous network layers.
The DT's ability to fuse multi-domain data sources, including weather forecasts, energy availability, and mobility trajectories, makes it particularly well-suited for orchestrating ISATNs, which exhibit high variability in traffic patterns and environmental conditions. Through closed-loop optimization, the DT can implement adaptive configurations in real time, dynamically aligning resource allocation with both performance and carbon reduction goals. This level of intelligence is critical to meeting net-zero objectives without compromising user experience, establishing DTs as a foundational component of future 6G network design. By embedding sustainability metrics such as grams of CO$_2$-equivalent per delivered bit (gCO$_2$e/bit) into network decision-making, DT-enabled orchestration offers a clear pathway toward achieving energy-efficient, low-carbon communications at a global scale.

\section{Architecture for Carbon-Optimal ISATN}

\begin{figure}[t]
  \centering
  \includegraphics[width=0.48\textwidth]{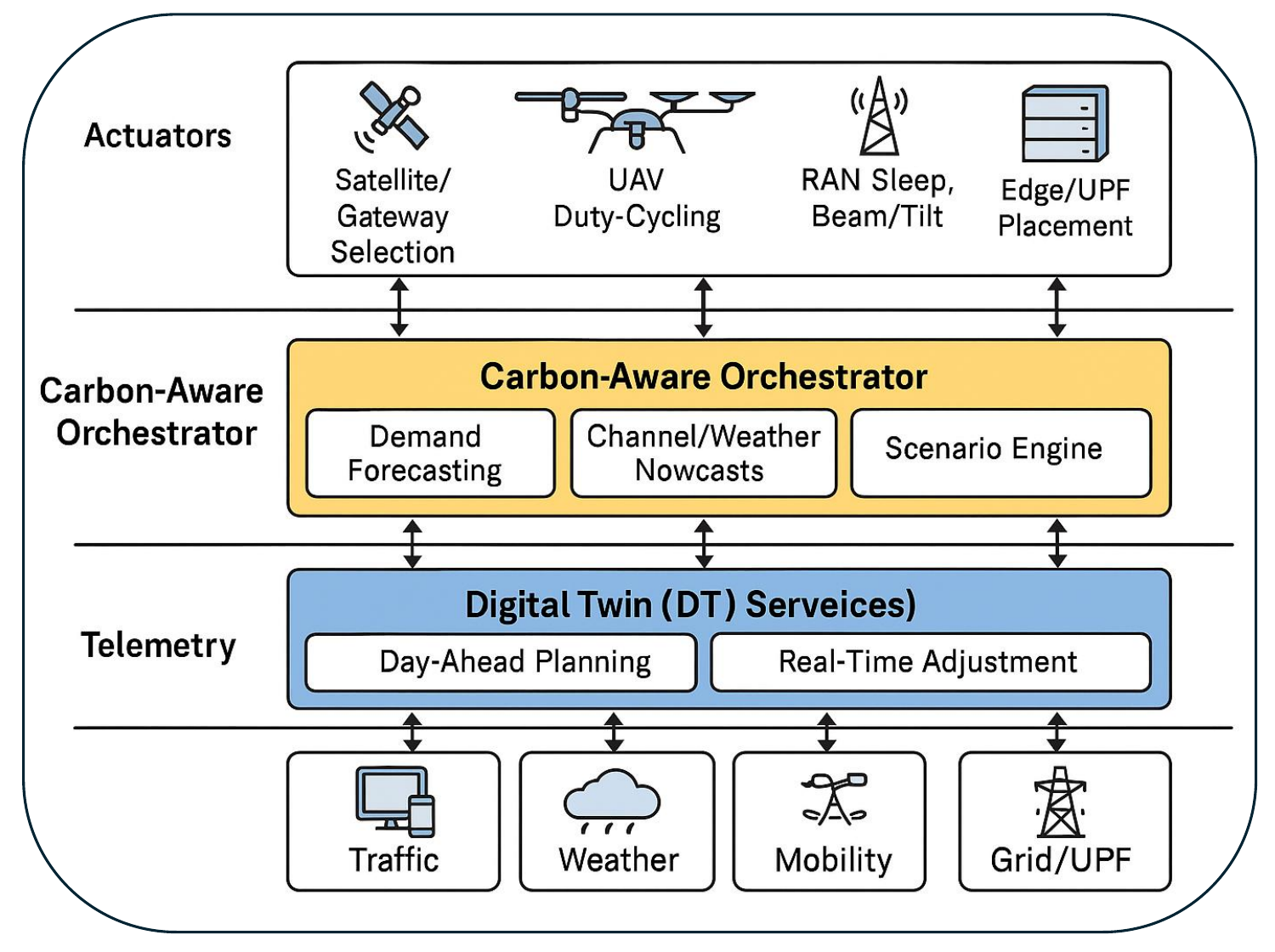}
  \caption{Carbon-aware orchestration framework for ISATNs. Telemetry informs DT planning and adjustment, the orchestrator generates policies, and actuators execute them, aligning with the PDCA cycle (i.e., Plan, Do, Check, and Act).}
  \label{fig:framework}
\end{figure}

This section presents the architectural framework for carbon-aware orchestration in ISATNs. The system is structured into four core components, including telemetry sources, DT services, an orchestrator, and actuators as shown in Fig. \ref{fig:framework}. Together, these components implement a PDCA control loop that continuously monitors network conditions, forecasts demand and energy trends, evaluates operational outcomes, and re-optimizes policies in real time. This closed-loop design offers a scalable and adaptive foundation for achieving sustainability objectives in 6G networks while ensuring high QoS.

\subsection{Telemetry Sources}
Efficient carbon-aware orchestration relies on a comprehensive telemetry framework that provides multi-dimensional visibility into both network performance and environmental factors. Traffic telemetry is collected at fine temporal resolutions (e.g., sub-second granularity) through 3GPP-compliant network management interfaces, enabling precise load characterization and user demand profiling. Weather telemetry, including rain fade, atmospheric attenuation, and temperature data, is gathered from satellite-based sensors and terrestrial weather stations, directly influencing link adaptation and beam steering decisions. Mobility telemetry tracks the trajectories of UAV platforms, satellite orbital parameters, and user mobility patterns using GPS, inertial sensors, and predictive analytics, while grid telemetry integrates real-time and forecasted carbon intensity data sourced from energy providers and market signals. Together, these telemetry streams create a high-fidelity view of the operational landscape, forming the foundation for predictive modeling, simulation, and real-time decision-making within the DT framework \cite{zilberman2023toward}.

\subsection{Digital Twin (DT) Services}
The DT layer serves as the intelligence core of the architecture, transforming telemetry into actionable insights that guide orchestration decisions. The DT maintains a high-fidelity, continuously updated replica of the physical network, capturing dynamic elements such as topology changes, traffic distributions, mobility patterns, and energy supply conditions. Advanced forecasting modules leverage machine learning models, such as recurrent neural networks or graph neural networks, to predict traffic load at fine spatial and temporal resolutions. Channel and weather nowcasting services incorporate environmental data, including rain fade models and satellite propagation characteristics, to proactively identify potential link degradation and guide adaptive resource allocation. A dedicated scenario engine enables the simulation of ``what-if'' events, such as sudden demand spikes, energy shortfalls, or extreme weather conditions, allowing operators to evaluate trade-offs between carbon intensity and QoS. Through its integration of prediction, simulation, and real-time data assimilation, the DT layer acts as a decision-support system, providing the orchestrator with precise recommendations for energy- and carbon-efficient operations.

\subsection{Carbon-Aware Orchestrator}
The orchestrator is responsible for multi-timescale decision-making, implementing optimization strategies that balance sustainability objectives with performance guarantees. A day-ahead planning module uses forecasted traffic, weather conditions, and grid carbon intensity predictions to generate baseline operational schedules. These plans define satellite gateway assignments, UAV deployment strategies, and energy budgets that align with expected renewable energy availability. In parallel, a real-time adjustment loop continuously refines decisions in response to deviations from forecasts, enabling dynamic adaptation to unexpected demand fluctuations, environmental changes, or variations in grid carbon intensity. Optimization techniques can incorporate mixed-integer linear programming, Reinforcement Learning (RL), or distributed optimization frameworks, with carbon emissions explicitly modeled as a primary constraint alongside throughput, latency, and reliability metrics \cite{chi2024metric}. In the context of the PDCA control loop, the day-ahead module corresponds to the \textit{Plan} stage, while the real-time loop performs \textit{Do}, \textit{Check}, and \textit{Act} by executing policies, monitoring outcomes, and re-optimizing in response to changes. By integrating sustainability considerations at every decision layer, the orchestrator enables proactive, data-driven network control that dynamically minimizes gCO$_2$e/bit.


\subsection{Actuators}
The actuator layer enforces the orchestrator’s decisions, closing the loop between planning and execution. Key actuators include carbon-aware satellite and gateway selection mechanisms that optimize backhaul routing based on both link quality and grid carbon intensity at ground stations. UAV platforms are dynamically switched on or off, or placed in low-power standby states, to reduce propulsion and communication energy usage during periods of low demand. Radio Access network (RAN) components employ energy-saving features such as micro-sleep and deep-sleep modes, further minimizing idle power consumption. Beam steering and adaptive power control ensure that transmission resources are efficiently directed, while dynamic edge function placement aligns compute resources with traffic hotspots powered by low-carbon energy sources. Collectively, these actuators translate high-level orchestration policies into precise physical network adjustments, ensuring that the entire system remains responsive, efficient, and carbon-optimized.

\section{Orchestration Workflow}
The orchestration workflow integrates three key components, including day-ahead planning, real-time orchestration, and the carbon and energy model. Day-ahead planning establishes a proactive baseline schedule, while real-time orchestration ensures rapid adaptation to dynamic network conditions. As illustrated in Fig. 2, both layers operate in a closed loop with the DT, enabling carbon-aware and resilient decision-making in ISATNs.

\begin{figure}[htbp]
    \centering
    \includegraphics[width=0.9\linewidth]{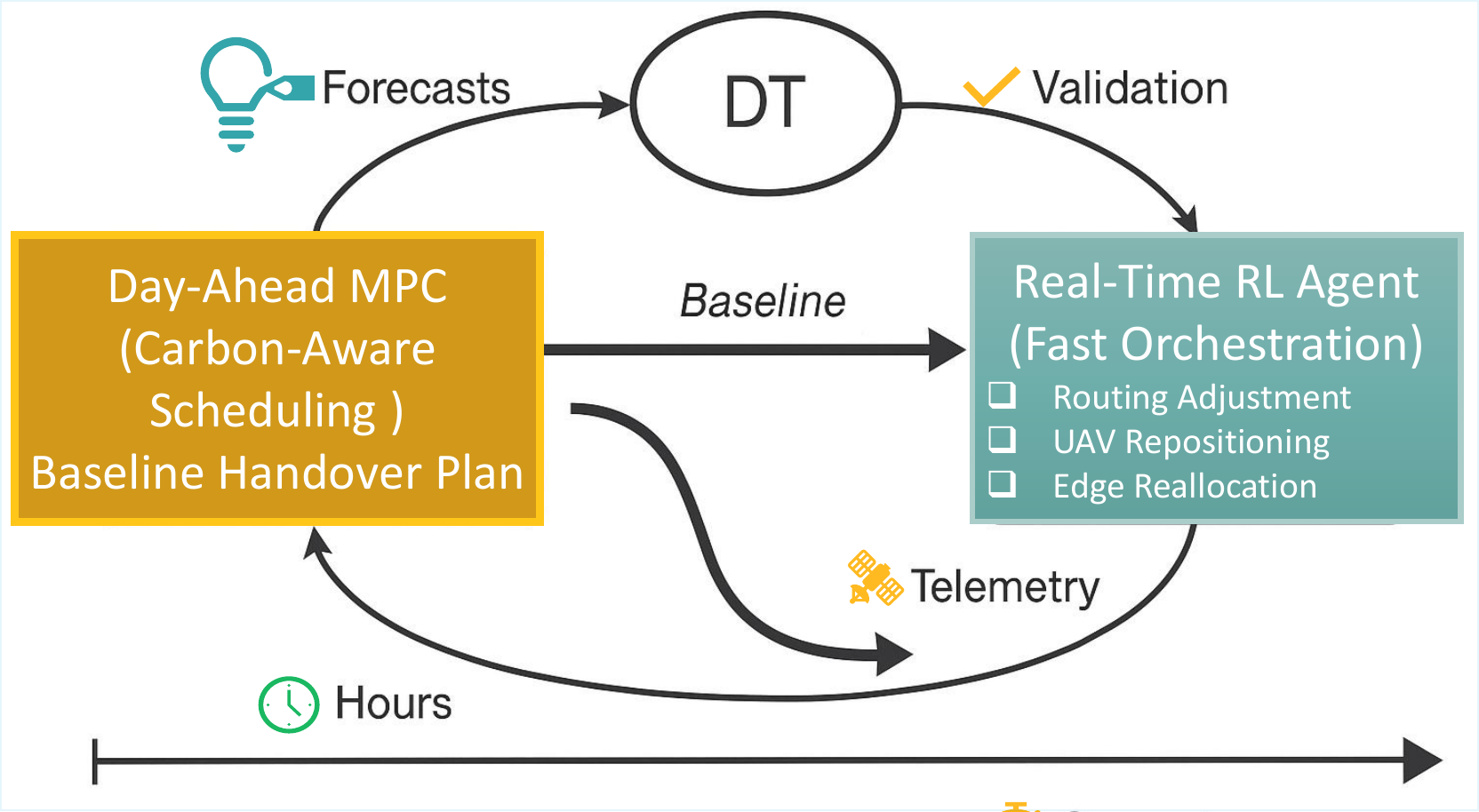}
    \caption{Dual-timescale carbon-aware orchestration in ISATNs.
    Day-ahead MPC generates a baseline plan over hours, while the real-time RL agent performs fast adjustments over seconds.
    Both are supported by the Digital Twin for forecasting, validation, and telemetry feedback.}
    \label{fig:dual_orchestration}
\end{figure}

\subsection{Day-Ahead Planning}
The day-ahead planning phase establishes a baseline operational schedule for the next cycle by leveraging predictive analytics and DT capabilities. This stage operates over a longer time horizon, typically 12–24 hours, allowing the orchestrator to make proactive decisions based on forecasts of network demand, channel quality, and energy availability. Traffic predictions are generated using historical telemetry and contextual information such as time of day, user mobility patterns, and application mix, while channel conditions are modeled using satellite ephemeris data, rain fade probabilities, and UAV trajectory forecasts. Grid carbon intensity forecasts are incorporated to identify periods of low-carbon energy availability, enabling the orchestrator to align compute-intensive workloads and network activities with cleaner energy sources.
In this stage, the DT simulates multiple candidate configurations that specify satellite handover schedules, UAV deployment and duty cycles, and base station sleep windows. Edge service placement is also optimized to ensure compute resources are strategically positioned to minimize latency and energy consumption. Each candidate is evaluated in terms of its estimated carbon footprint, QoS performance, and operational risk. By ranking these configurations, the orchestrator selects a baseline plan that achieves a balance between carbon reduction goals and user experience guarantees. This baseline plan serves as a stable foundation for real-time orchestration, reducing computational overhead and allowing the system to operate efficiently under anticipated conditions. The day-ahead planning phase is inspired by approaches used in energy markets, where load forecasts and carbon intensity predictions guide power dispatch. In ISATNs, this concept extends beyond energy allocation to include spatial and temporal orchestration of aerial platforms, satellite beams, and terrestrial resources. By integrating telemetry-driven forecasting, scenario analysis, and policy evaluation, the day-ahead stage ensures that the network enters the operational cycle with a carbon-aware strategy that minimizes the need for drastic interventions in real time \cite{you2022digital}.





\subsection{Real-Time Orchestration}
The real-time orchestration loop operates on short time scales, ranging from seconds to minutes, to complement the day-ahead planning phase. Its primary objective is to ensure that the network remains resilient to unpredictable events, such as sudden traffic surges, user mobility changes, weather-induced link degradation, or unforeseen fluctuations in grid carbon intensity. While the baseline schedule provides a stable operational foundation, real-time orchestration continuously adapts the plan to maintain both service quality and sustainability objectives under dynamic conditions.

This loop leverages continuous telemetry streams collected from satellites, UAV platforms, terrestrial base stations, and power systems. KPIs, including link quality, latency, throughput, node utilization, and carbon intensity, are monitored to detect deviations from expected behavior. When anomalies are identified, the orchestrator applies localized adjustments that minimize disruption and computational overhead. Corrective actions include dynamic routing to lower-carbon gateways \cite{struhar2024hierarchical}, reallocation of edge computing tasks to balance workload and energy efficiency, UAV repositioning or activation/deactivation to conserve propulsion energy, and fine-grained power adaptation in the RAN.
The decision-making process is supported by the DT, which maintains a near-real-time virtual representation of the network. The DT rapidly evaluates the potential impact of corrective actions through lightweight simulations, enabling the orchestrator to select minimally invasive adjustments that satisfy latency and availability requirements while reducing energy use and emissions. By adopting a minimal-change strategy, the network avoids oscillations and unnecessary state transitions, ensuring operational stability even in volatile environments.
This capability is particularly valuable for ISATNs, which face high variability in environmental conditions and traffic patterns, as well as heterogeneous energy and operational constraints across system elements. Real-time orchestration transforms the network into a self-optimizing, sustainability-aware system that complements long-term planning strategies with fine-grained, adaptive control.

\subsection{Carbon and Energy Model}
The carbon and energy model provides a foundation for evaluating the environmental and operational performance of the proposed orchestration framework. This model captures how energy is consumed across ISATN components and translates energy use into carbon impact using grid carbon intensity data.
Each network element, satellites, UAV platforms, terrestrial base stations, and edge computing nodes, has a distinct energy consumption profile influenced by transmission power, backhaul and processing demands, and idle or standby states. UAV platforms, for example, exhibit high propulsion energy costs that vary with mobility and operational altitude, while terrestrial base stations consume significant energy even when underutilized. Satellites add complexity due to dynamic beam steering and inter-satellite link operations, requiring careful modeling of energy distribution across the constellation \cite{zilberman2023toward}.

To assess environmental impact, real-time grid carbon intensity data is incorporated into the model. This allows the orchestration system to estimate the carbon footprint of network decisions based on the time and location of energy use. By aligning network activities with periods and regions of low-carbon energy availability, the orchestrator can significantly reduce emissions without compromising service quality.

This model also considers renewable energy availability and its variability across geographic regions, enabling the system to prioritize edge or gateway sites powered by renewable sources. The resulting carbon and energy framework supports decision-making at both day-ahead and real-time scales, guiding actions such as UAV scheduling, base station sleep cycles, and workload placement. Together, these features ensure that network operations are optimized holistically for both energy efficiency and carbon reduction, providing the necessary feedback to evaluate and refine orchestration strategies \cite{patel2024modeling}.

\section{Solution Approach}


The proposed orchestration strategy integrates two complementary decision-making mechanisms operating at different time scales. The first is a Model Predictive Control (MPC)–based day-ahead scheduler for proactive planning, and the second is a RL–based real-time controller for fine-grained adjustments, both designed to enable the DT to generate low-carbon, high-efficiency network configurations that remain robust under dynamic operating conditions.

\subsection{Day-Ahead MPC (Carbon-Aware Scheduling)}
The day-ahead component of the framework uses model predictive control to optimize orchestration decisions over a multi-hour time horizon. The DT forecasts traffic demand, weather conditions, and grid carbon intensity at sub-hourly intervals using telemetry data and historical patterns. These predictions allow the orchestrator to plan satellite handover schedules, UAV deployment strategies, and base station sleep cycles in advance.
Candidate configurations are generated and evaluated in terms of expected carbon footprint, network energy consumption, and QoS performance. To reduce computational complexity, a convex or Mixed-Integer Quadratic Programming (MIQP) relaxation is applied, producing a baseline orchestration plan that aligns network activity with anticipated renewable energy availability and low-carbon time windows. This baseline plan serves as a reference for real-time adjustments and reduces the need for frequent large-scale re-optimization \cite{perin2022ease}.

\subsection{Online RL (Fast Orchestration)}
While day-ahead MPC provides a stable baseline, network dynamics such as unexpected traffic spikes, weather disruptions, or sudden changes in grid carbon intensity require fast, localized decisions. To address this, we integrate an RL agent operating in real time. The RL module is implemented as an actor–critic agent that continuously interacts with the DT, receiving telemetry-derived KPIs (e.g., link quality, latency, throughput, carbon metrics) and selecting corrective actions.

Actions include adaptive traffic rerouting, UAV repositioning or activation/deactivation, dynamic beam steering, and edge resource reallocation. The RL agent is trained to minimize operational carbon per bit while satisfying strict service constraints. During inference, the agent reacts within seconds, enabling fine-grained orchestration without incurring the overhead of solving optimization problems from scratch \cite{rao2024dar}.

This hybrid approach leverages the strengths of both techniques, MPC provides long-term, carbon-aware scheduling with global network context, while RL ensures rapid responsiveness to short-term disturbances. Together, they create a closed-loop control framework in which the DT continuously validates and improves orchestration policies, ensuring scalability and sustainability in ISATNs.



\section{Results and Discussion}
This section summarizes the simulation environment, network assumptions, and evaluation methodology, followed by key results demonstrating the effectiveness of the proposed orchestration strategy.

\subsection{Simulation Setup}

We evaluate the proposed orchestration framework over a 7-day trace (hourly resolution) in a metropolitan region spanning urban, suburban, and rural zones. The ISATN scenario comprises:

\begin{itemize}
    \item \textbf{Satellite segment:} A 72-satellite LEO constellation (six planes, 600\,km altitude) with inter-satellite links.
    \item \textbf{Terrestrial RAN:} 180 sites (60 macro and 120 small cells) with multi-level sleep modes.
    \item \textbf{Aerial relays:} 24 UAVs (4\,h endurance, 15\,km coverage) supported by six rapid battery swap pads.
    \item \textbf{Edge resources:} Eight gateways with computing resources for local processing and caching.
\end{itemize}

Traffic demand includes three categories:
\begin{itemize}
    \item \textbf{Enhanced Mobile Broadband (eMBB):} Diurnal peaks with dense urban concentration.
    \item \textbf{Ultra-Reliable Low-Latency Communications (URLLC):} Industrial and emergency services, with a $<5$\,ms latency target.
    \item \textbf{Massive IoT:} Bursty traffic dominated by rural sensing.
\end{itemize}

Mobility reflects commuter flows between suburban and urban areas, along with random waypoint mobility in rural zones. Wireless propagation is modeled using ITU-R P.618 rain attenuation and 3GPP TR 38.901 path loss models \cite{itur_p618,3gpp_38901}. Two rain events (Day~3 evening and Day~6 afternoon) degrade Ka-band satellite and microwave backhaul links. Adaptive modulation, coding, power control, and beam steering are applied across LEO satellites and macro cells.

Carbon intensity and renewable generation traces are based on CAISO historical profiles \cite{caiso}, with coastal regions reaching midday solar-driven minima of 45–65\% renewables, while inland regions experience evening peaks of 20–35\%. Device power models include macro RUs/BBUs (1.8–2.6,kW, active), small cells (70–110,W, active), edge servers (0.6–1.0,kW), and UAV propulsion (300–500,W during cruise and 700–900,W while hovering). Telemetry for network KPIs is sampled every 5–30,s, but orchestration decisions are executed at hourly resolution, consistent with carbon intensity signals.

These models and traces provide the foundation for evaluation in the next section, where we compare the proposed orchestration with baseline strategies under realistic traffic, weather, and energy conditions.

\subsection{Results}
We evaluate our orchestration strategy on a 7-day trace (hourly resolution) using real CAISO grid carbon intensity data, which exhibits strong diurnal variability with midday solar dips and evening peaks. The simulation covers urban, suburban, and rural zones, and includes two rain events (Day~3 evening and Day~6 afternoon) to stress-test performance. We compare four strategies: a QoS-focused orchestration optimized for throughput and latency, an energy-minimization heuristic that ignores carbon variation, a static configuration with fixed routes and assignments, and the proposed MPC+RL orchestration validated by a DT.

\subsubsection{Carbon Reduction}
\begin{figure}[t]
  \centering
  \includegraphics[width=0.48\textwidth]{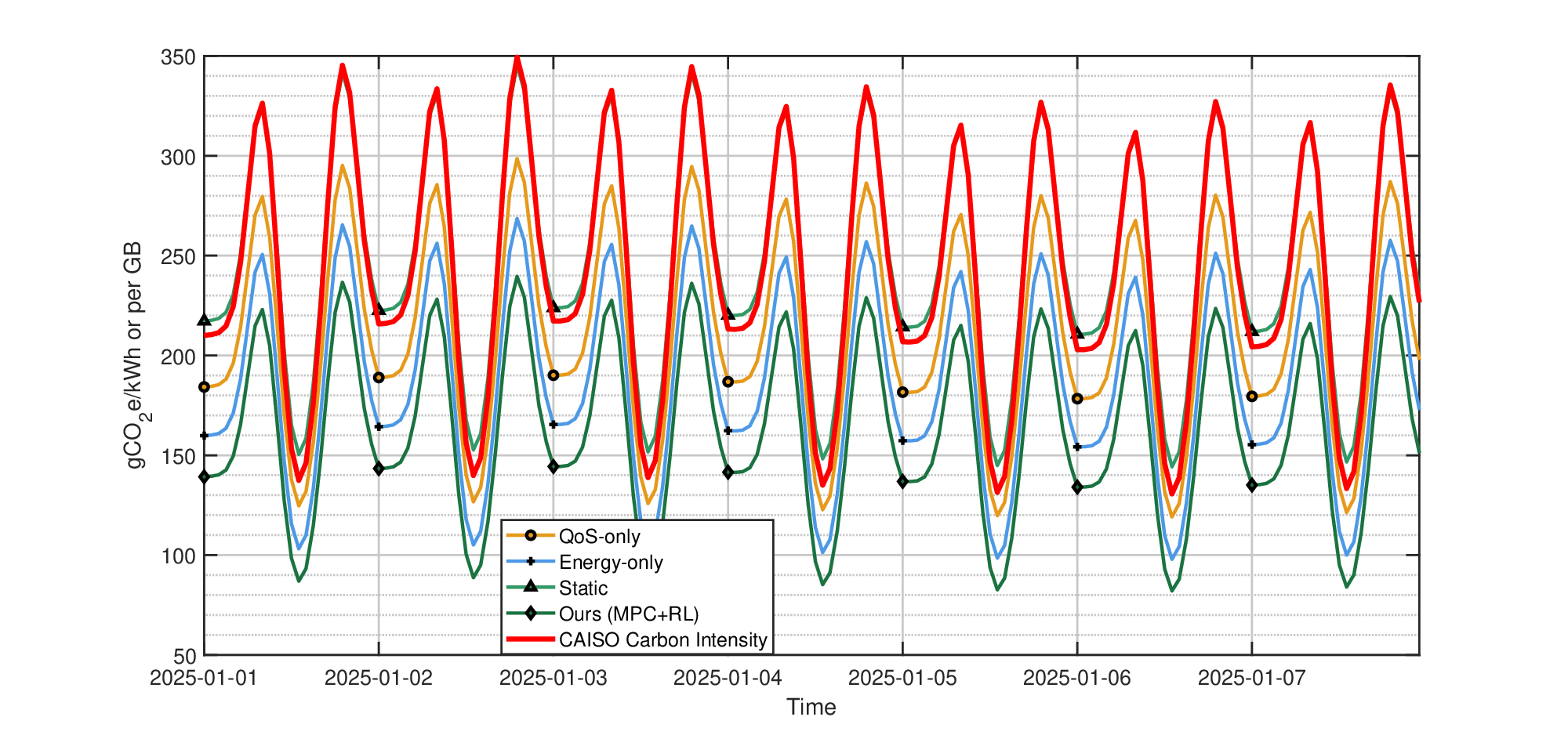}
  \caption{Carbon intensity over 7 days with emissions per GB for different strategies.
  The proposed MPC+RL orchestration closely aligns traffic with low-carbon periods.}
  \label{fig:carbon_trace}
\end{figure}

Fig.~\ref{fig:carbon_trace} shows that the proposed orchestration achieves a 29\% reduction in gCO$_2$/GB compared to QoS-only optimization and a 38\% reduction relative to the static configuration. Energy-only orchestration lowers total energy use but fails to account for carbon timing, limiting emissions savings to 18\%. In contrast, MPC+RL scheduling leverages daily solar-rich periods and low-carbon routing to achieve the lowest overall emissions, closely tracking real CAISO carbon intensity trends.

Fig.~\ref{fig:carbon_trace} shows the gCO$_2$e/GB trajectory over the 7-day horizon. The proposed orchestration achieves a 29\% reduction compared to QoS-only optimization and a 38\% reduction relative to the static configuration. The energy-only scheme reduces overall energy consumption but fails to align with low-carbon periods, limiting its savings to 18\%. In contrast, MPC+RL scheduling adapts to periods of low carbon intensity (i.e.,  primarily driven by midday solar generation) and leverages low-carbon routing opportunities, thus achieving the lowest overall emissions while closely tracking CAISO carbon intensity trends.

\subsubsection{Energy Efficiency and Renewable Utilization}
\begin{figure}[t]
  \centering
  \includegraphics[width=0.48\textwidth]{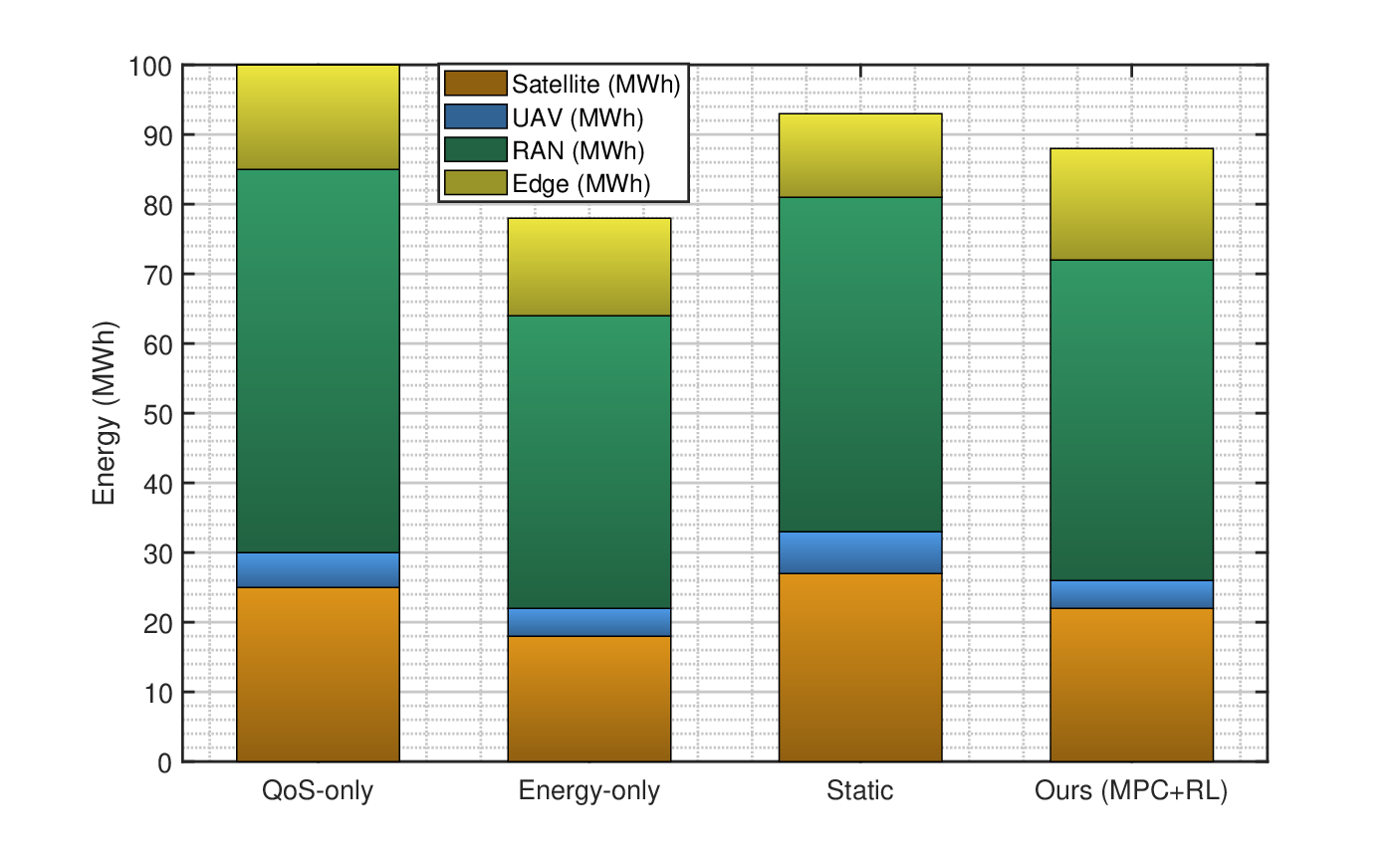}
  \caption{Weekly energy consumption by network layer. MPC+RL orchestration lowers RAN idle and satellite energy while shifting workloads to renewable-rich nodes.}
  \label{fig:energy_breakdown}
\end{figure}

Our approach balances emissions reduction with efficiency. Figure~\ref{fig:energy_breakdown} shows weekly energy consumption by layer, demonstrating clear reductions in idle RAN and satellite link energy, with a slight increase in edge compute energy due to workload shifting. Overall, total energy use decreases by about 12\% relative to QoS-only orchestration, while renewable energy utilization (measured separately) rises from 43\% to 58\% through time- and location-aware routing. These gains are enabled by ISATN-specific control knobs, including handovers to cleaner coastal gateways, UAV duty-cycling to reduce propulsion overhead, and shifting workloads toward renewable-rich edge sites. In combination, these mechanisms lower total consumption and enhance renewable alignment, complementing the carbon-intensity reductions discussed earlier.

\subsubsection{Resilience to Rain Events and Traffic Surges}
\begin{figure}[t]
  \centering
  \includegraphics[width=0.48\textwidth]{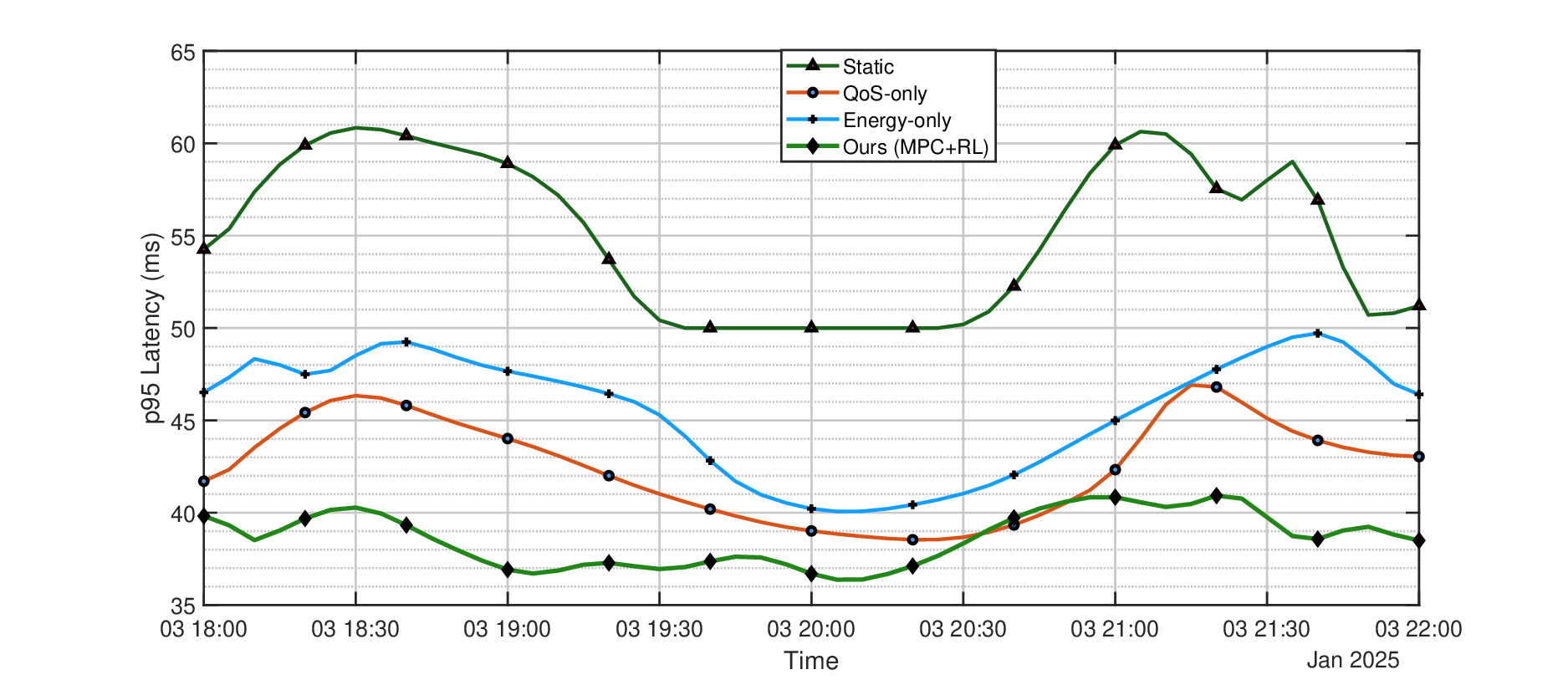}
  \caption{p95 latency during the Day~3 rain event. MPC+RL orchestration stabilizes within 60 seconds, outperforming other strategies.}
  \label{fig:latency_rain}
\end{figure}

During rain fade and traffic surges, MPC+RL adapts quickly using RL-driven route reallocation, beam steering, and edge placement. Figure~\ref{fig:latency_rain} shows that our method achieves about 10\% lower p95 latency than QoS-only optimization and maintains stability within 60 seconds, whereas QoS-only requires several minutes to recover. Compared to static policies, MPC+RL also significantly reduces SLA violations (measured separately) by approximately 40\%, demonstrating resilience under highly dynamic channel and traffic conditions.

These results demonstrate that carbon-aware orchestration requires more than minimizing energy. By leveraging a DT for predictive planning (MPC) and low-overhead corrections (RL), our system reduces emissions, improves renewable energy usage, and maintains QoS even under adverse conditions.

\section{Open Research Challenges}
Despite the promise of carbon-aware orchestration in ISATNs, several open challenges remain to be addressed:
\subsection{Balancing Digital Twin fidelity and overhead} DT technology enables ``what-if'' analyses to evaluate the carbon and performance implications of different orchestration strategies prior to real-world implementation. However, a persistent challenge lies in balancing fidelity and overhead. High-fidelity DTs that capture fine-grained dynamics, such as UAV mobility patterns, interference conditions, or renewable energy availability, can provide accurate insights but demand substantial computational resources, frequent data synchronization, and significant communication bandwidth. In contrast, low-fidelity DTs reduce overhead but risk producing oversimplified or misleading results. This trade-off complicates the validation of what-if analyses, as conclusions may vary depending on the level of detail maintained within the DT. To address these challenges, future research should prioritize the development of adaptive DT architectures capable of dynamically adjusting their level of fidelity according to operational requirements. Promising directions also include hybrid validation methods that combine lightweight simulations with selective real-world measurements, thus reducing dependence on exhaustive data synchronization while maintaining reliability. Furthermore, standardized benchmarking frameworks are needed to systematically quantify the trade-offs between fidelity, overhead, and carbon efficiency, ensuring that what-if analyses remain both accurate and sustainable.
\subsection{Trade-offs between embodied and operational emissions}
While most studies on carbon-aware orchestration in ISATNs emphasize operational emissions, such as energy consumed during communication, computation, and mobility, embodied emissions remain largely overlooked. Embodied emissions are the carbon costs incurred throughout the lifecycle of network components, including raw material extraction, manufacturing, deployment, and disposal. Emerging hardware technologies such as Reconfigurable Intelligent Surfaces (RIS) and UAV platforms exemplify this trade-off. For example, RIS panels may reduce operational energy consumption by enhancing propagation efficiency. However, their large-scale fabrication and limited recycling pathways introduce significant embodied emissions. Similarly, UAVs designed for extended endurance or heavy payloads require energy-intensive manufacturing processes and frequent battery replacements, amplifying lifecycle emissions. Evaluating sustainability solely based on operational efficiency therefore risks underestimating the true environmental footprint of ISATNs.

Future work should adopt a holistic perspective that jointly accounts for embodied and operational emissions. Lifecycle assessment methodologies tailored to ISATN components could provide a standardized framework for quantifying these impacts. Another promising direction is the co-design of hardware and orchestration policies, for instance, integrating RIS only when the operational energy savings outweigh their embodied carbon costs, or designing UAV platforms with modular, recyclable components to minimize lifecycle impacts. In addition, carbon-aware orchestration algorithms should incorporate embodied emission budgets into their optimization objectives, ensuring that short-term operational gains do not compromise long-term sustainability.

\subsection{Inter-domain coordination}
Carbon-aware orchestration in ISATNs cannot be optimized within a single technological domain. Instead, performance and sustainability depend on effective coordination across multiple domains, including spectrum management, gateway operations, and power grid interactions. Spectrum allocation, for instance, directly affects interference levels and energy efficiency but is often fragmented across regulatory jurisdictions. Similarly, gateways that interconnect satellites, aerial platforms, and terrestrial nodes must balance data throughput, latency, and power consumption while adapting to variable renewable energy availability. At the same time, tight coupling with electrical grids introduces additional complexity, as renewable energy generation is intermittent and geographically heterogeneous. A lack of integrated control mechanisms across these domains risks creating local optimizations that increase global inefficiencies, undermining both carbon reduction and service quality objectives.

Future research should focus on multi-domain orchestration frameworks that jointly optimize communication, computation, and energy flows across ISATNs. One promising direction is the development of cross-layer protocols that dynamically adapt spectrum usage while considering gateway workloads and grid carbon intensity signals. Another is the design of interoperable control architectures that allow satellites, UAVs, and terrestrial nodes to share information about spectrum occupancy, energy availability, and traffic demands in near real time. Incorporating grid-aware decision-making into orchestration algorithms, for example, scheduling energy-intensive operations when renewable availability is high, could further enhance sustainability.

\section{Conclusion}
This paper has presented a carbon-aware orchestration framework for Integrated Satellite–Aerial–Terrestrial Networks (ISATNs) leveraging Digital Twin (DT) technology. By introducing grams of CO$_2$-equivalent per bit (gCO$_2$/bit) as a primary sustainability metric, the proposed framework explicitly incorporates environmental objectives alongside QoS considerations such as throughput, latency, and availability. The orchestration design integrates a multi-timescale PDCA control loop, where day-ahead forecasting establishes a proactive baseline and real-time RL enables fine-grained adaptation under dynamic conditions.  The framework leverages ISATN-specific controls—carbon-aware handovers, UAV duty-cycling, and renewable-aware edge placement—to cut gCO$_2$/bit by up to 29\% versus QoS-only orchestration, while enhancing renewable use and resilience By embedding carbon-awareness into orchestration, this work shows that sustainability objectives can be achieved without compromising QoS, reinforcing the role of ISATNs in green 6G. Future research will extend the framework toward multi-domain interoperability, life-cycle carbon accounting of hardware platforms, and integration with emerging paradigms such as semantic communications and AI-native networks.

\bibliographystyle{IEEEtran}
\bibliography{refs} 

\end{document}